\documentclass[10pt]{IEEEtran}

\usepackage{amsmath,amsthm,amssymb}

\usepackage{float}
\usepackage{wrapfig}

\usepackage[utf8x]{inputenc}
\usepackage{epsfig}
\usepackage{rotating}
\usepackage{ucs}
\usepackage[utf8x]{inputenc}
\usepackage{etex}
\usepackage{overpic}
\usepackage{listings}
\usepackage{amsthm}
\usepackage{amsmath,amssymb,nopageno}
\usepackage{subfigure}
\usepackage{color}
\usepackage[all]{xy}
\usepackage{float}

\usepackage{amssymb}
\usepackage{cite}
\usepackage{url}
\usepackage{enumitem}
\renewcommand{\P}{{\mathbb P}}
\newcommand{\R}{\mathbb{R}}

\DeclareMathOperator*{\minimize}{minimize}

\begin{document}

\title{On DICE-free Smart Cities, Particulate Matter, and Feedback-Enabled Access Control\thanks{This paper has been submitted to the IEEE Access under the title \emph{Distributed ledger enabled control of tyre induced particulate matter}}}
\author{
    \IEEEauthorblockN{}
       Panagiota~Katsikouli, Pietro Ferraro, David Timoney, Marc Masen\protect\footnote{Mechanical Engineering, Imperial College London}, Robert Shorten\\
     \IEEEauthorblockA{
        }}
\maketitle
\begin{abstract}
The link between transport related emissions and human health is a major issue for city municipalities worldwide. PM emissions from exhaust and non-exhaust sources are one of the main worrying contributors to air-pollution. In this paper, we challenge the notion that a ban on internal combustion engine vehicles will result in clean and safe air in our cities, since emissions from tyres and other non-exhaust sources are expected to increase in the near future. To this end, we present data from the city of Dublin that document that the current amount of tyre-related PM emissions in the city might already be above or close to the levels deemed safe by the World Health Organization. As a solution to this problem, we present a feedback-enabled distributed access control mechanism and ride-sharing scheme to limit the number of vehicles in a city and therefore maintain the amount of transport-related PM to safe levels. 
\end{abstract}

\section{Introduction}
The link between transport related emissions, and human health, is a major issue for city municipalities worldwide. Diesel and other internal combustion engine based motor vehicles are considered to be the major culprit in this regard as they are associated with the generation of a number of harmful emissions. Apart from the link to global warming through the generation of carbon-dioxide, such vehicles are also known to produce other airborne pollutants such as nitrogen oxides, ozone, benzene, carbon monoxide, and particulate matter (PM) of varying size, all of which are considered harmful to human health. In a recent global review~\cite{chest_review}, it is stated that air pollution, in general, could be damaging every organ and every cell in the human body, showing a potential link between toxic air and skin damages, fertility, asthma and allergies to children and adults.\newline 

One of the main reasons behind all this is considered to be PM emissions. PM is a generic term used for a type of pollutants that consist of a complex and varied mix of particles suspended in air. Among all the airborne pollutants PM is particularly worrying due to its ability to enter the bloodstream and reach major organs in the human body. There is rich literature documenting the link between PM and its effects on human health~\cite{valavanidis,gehring,eea2014,harvard_2006,air_survey_2014,pm_brazil_2011}. In particular, the World Health Organization reports that ``adverse health effects of PM are due to exposure over both short (hours, days) and long (months, years) terms and include respiratory and cardiovascular morbidity (aggravation of asthma, respiratory symptoms, increase in hospital admissions), as well as mortality from cardiovascular and respiratory diseases and from lung cancer''~\cite{who}. Smaller PM particles tend to be more harmful to humans compared to larger ones, as they can travel deeper into the respiratory system~\cite{who,eea2014}. Some of the health effects related to PM include oxidative stress, inflammation and early atherosclerosis. Other studies have shown that smaller particles may go into the bloodstream and thus translocate to the liver, the kidneys or the brain (see~\cite{non_exhaust} and references within). Transport related emissions are a significant contributor to airborne PM levels that harm our health. In a recent study~\cite{dementia}, it is shown that living near major roads (i.e. near emissions from vehicles) is associated with increased risk of dementia. The reduction of air quality and population exposure to harmful pollutants as a result of road passenger transportation is discussed in~\cite{toronto_2010}, in a case study in the Greater Toronto Area. The survey described in~\cite{air_survey_2014} focuses on air pollution originating from non-exhaust emissions such as brake and tyre wear, and highlights the related impact to human health as well as the significance of particulate matter reduction.\newline

Roughly speaking, three avenues are being explored worldwide in the fight against urban pollution:
\begin{itemize}
\item[(i)] outright bans on polluting vehicles and embracing zero tailpipe emission vehicles in certain city zones;
\item[(ii)] measuring air quality as a means to better informing citizens of zones of higher pollution \cite{francesco}; and
\item[(iii)] developing smart mobility devices that seek to minimise the effect of polluting devices on citizens as they transport goods and individuals in our cities \cite{annie,hermann,shaun}.  
\end{itemize}
Option (i) whereby ultra-low emission zones are created by banning internal combustion engine (ICE) based vehicles in certain areas, in addition to embracing electric vehicles (EV's), has gained much traction worldwide and is being proposed for adoption in cities such as London and Dublin. Apart from the reduced tailpipe pollutants, an additional attraction of the switch from ICE to EV, is that it is beneficial from the perspective of global warming (reduced carbon dioxide), provided that the energy delivered to the EV's can be sourced in a green manner. Thus, reducing our dependency of ICE based vehicles would appear to be very beneficial; not only does the strategy achieve cleaner air but we also potentially tackle climate change through reduced production of carbon dioxide.
A major objective of this paper is to challenge the current focus on tailpipe emissions. To avoid any {\em misunderstanding}, we wholeheartedly endorse a reduced dependency on ICE based vehicles. However, the contemporary narrative is based on tailpipe emissions only, and while 
it is indeed true that EV's are zero tailpipe emission vehicles, the tailpipe is only one source of PM. Thus replacing one type of vehicle fleet with another type of vehicle fleet may not result in cities with safe levels of air quality, especially if the non-tailpipe sources of PM are significant. One contribution of this work is to use data from Dublin to argue that PM levels from tyres alone may be above that which is deemed safe by the World Health Organisation (WHO). While we are by no means the first to argue that tyres are an important source of PM (see in particular the excellent report~\cite{microplastics}), we strongly believe it is important that stake-holders be reminded of this message, particularly in the context of the current transitioning from ICE-vehicles towards an electrification of the vehicle fleet. The second part of our paper describes a distributed access control mechanism that both regulates tyre-based PM generation, and provides fair access to a city zone for a set of competing vehicles.\newline

The paper is organised as follows. In Section~\ref{sec:pm_emissions} we discuss tyre-wear related PM emissions, supported by numbers from various sources, with a focus on Dublin in Ireland. We review traffic produced air-pollution mitigation measures in Section~\ref{sec:related}. Our Access Control Mechanism is presented in detail in Section~\ref{sec:acs}. We simulate our system and present results in Section~\ref{sec:simulations} and conclude the paper in Section~\ref{sec:conclusion}.

\section{Elementary calculations}\label{sec:pm_emissions}
{\bf Particulate Matter: } PM is the product of brake and tyre wear from vehicles as well as a by-product of the engine combustion process. The most common classification of particulate matter is according to size: $P\!M_{10}$ for particles with at most $10\ \mu m$ diameter, $P\!M_{2.5}$ for particles with at most $2.5\ \mu m$ diameter, and ultrafine particles which have a diameter of less than 0.1 micrometres.
Smaller PM particles tend to
be more harmful compared to larger ones as they are able to get deeper into the respiratory system
with ultrafine particles being able to get into the bloodstream and therefore translocate into vital
organs such as the liver, the kidneys and the brain.\newline 

{\bf Designated safe levels of PM: } According to the WHO, for $P\!M_{2.5}$, the daily maximum deemed safe level on average is $25\ \mu g/m^3$, whereas the annual maximum permitted level is on average $10\ \mu g/m^3$. For $P\!M_{10}$, the maximum permitted levels are on average $50\ \mu g/m^3$ and $20\ \mu g/m^3$ on a daily and annual basis, respectively. In general, non-exhaust emissions (including brake and tyre wear, road surface wear and resuspension of road dust) resulting from road traffic, account for over 90\% of $P\!M_{10}$ and over 85\% of $P\!M{2.5}$ emissions from traffic~\cite{non_exhaust_electric}.\newline  

{\bf Approximate guess of airborne PM in Dublin:} To parse these numbers in terms of tyre abrasion for a city with a high volume of cars,  we note that in 2014, nearly 28,000 tonnes of tyre waste was managed in Ireland \cite{waste_tyres_stats}. Using publicly available data from the Central Statistics Office~\cite{cso_data} in Ireland, in 2018 approximately 540,000 private cars were continuously active in Dublin throughout the year with an average distance travelled of approximately 15,000 km per vehicle. We assume that approximately 1/3 of the vehicles (i.e. c. 170,000 vehicles) will change their tyres in a year in Dublin\footnote{A tyre is changed when it has reached a tread wear of approximately 2 mm (or 1.6 mm as is the legal minimum). This translates into approximately 35,000 km of travelled distance per vehicle; however, depending on the driving conditions, the travelled distance before a tyre is changed can vary from 10,000 km (harsh braking and acceleration, constant change of gears) to 80,000 km (perfect driving conditions and favourable road and weather).}. Depending on the type of the tyre and the road conditions, a vehicle (i.e, 4 tyres) loses $50-240 \  mg/km$ in mass~\cite{grigoratos}, which accounts for 4-6 kg of tyre mass lost before tyres are changed. By considering 4 kg of tyre mass lost per vehicle, we estimate that in Dublin, in 2018, at least 680,000 kg of tyre mass was wasted, 10\% of which goes airborne~\cite{microplastics,grigoratos} as PM. That corresponds to approximately 68,000 kg of particulate matter in a year, or 185 kg per day, in the city of Dublin. \cite{grigoratos} states that approximately 50\% of the $P\!M_{10}$ emissions (not specifically to air) fall in the $P\!M_{2.5}$ category. \cite{fausser} reports that c.90\% of airborne tyre wear particles are smaller than 1 micrometer in diameter (that is, in the $P\!M_{2.5}$ category). In~\cite{non_exhaust}, references of previous studies state that 3-7\% of tyre wear particles contribute to airborne $P\!M_{2.5}$.\newline 



\begin{figure}
\begin{center}
\includegraphics[scale=.3]{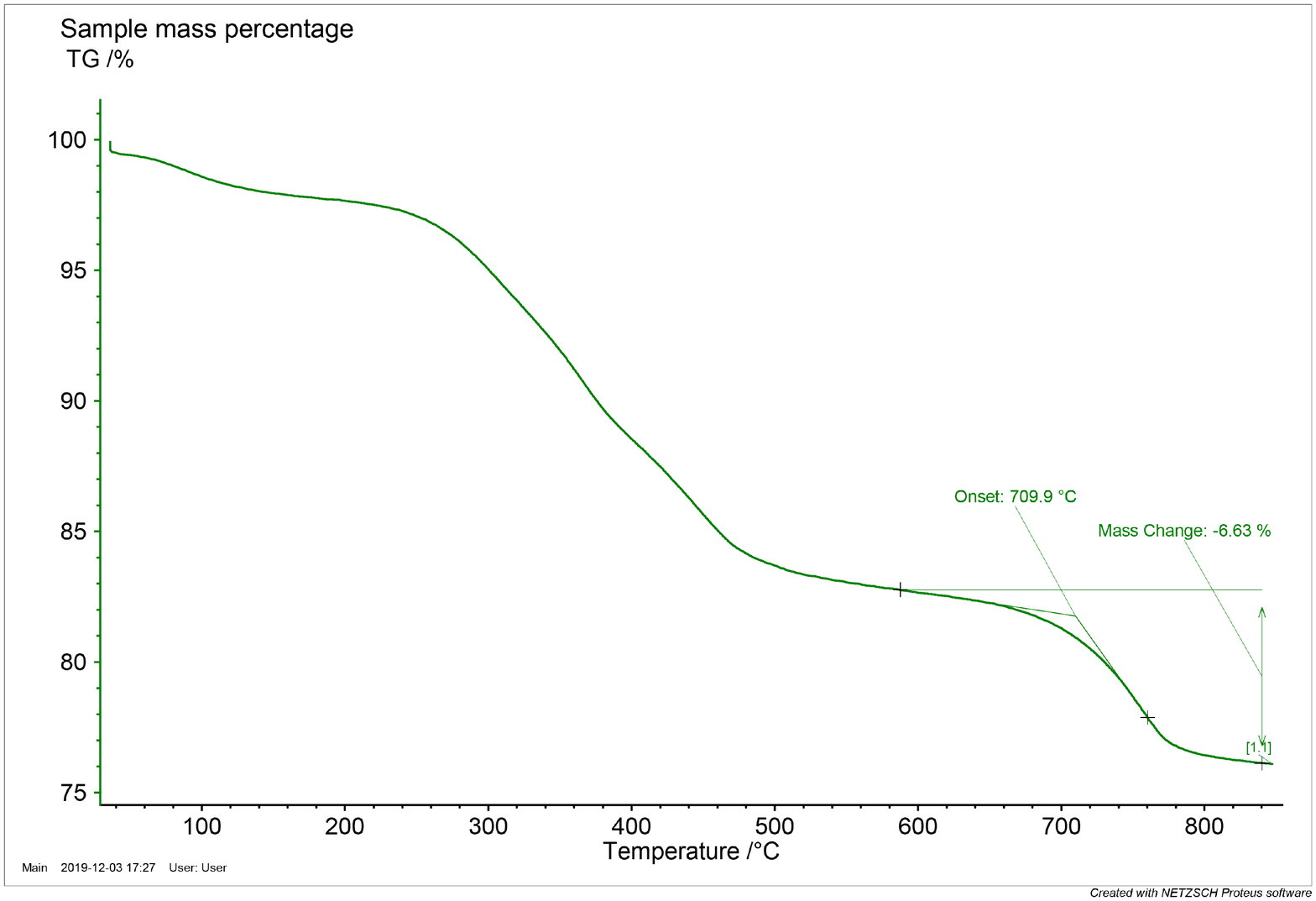}
\label{fig:tga}
\caption{Thermogravimetric analysis of road dust aggregated over all sites.}
\end{center}
\end{figure}

{\bf Rubber in road dust:} Finally, to determine the presence of tyre residuals in road dust in an urban environment, and thereby provide an experimental estimate of the relative proportion of rubber vs. non-rubber particles, road dust samples were collected from seven locations in central London: Brechin Place, Jay Mews, Exhibition Road, Bayswater Road, by Queensgate Station, Cromwell Road and Westway Roadprotect\footnote{Data collected at Imperial College London by S. Anderson, M . Mallya, H. Richardson, C. Ching.}. These locations include major thoroughfares as well as residential streets and an 'average' urban road dust sample was subsequently created by mixing these various collected samples. In order to measure the amount of tyre dust we employed thermogravimetric analysis \cite{Coats}, to measure mass change with temperature increase. The results of this analysis on the aggregated road sample are depicted in Figure \ref{fig:tga}. Tyre rubber ignites at approximately $ 700^{\circ} C$, and at this temperature the thermogravimetric curve shows a drop of $6.6\%$. This indicates that the mass fraction of rubber (or a similar material) in the collected road dust is approximately $6.6\%$.

\section{Air pollution mitigation measures}\label{sec:related}

Among the early solutions to reduce traffic related air-pollution has been the application of non-thermal plasma to diesel cars~\cite{non-thermal_2003}. Similar solutions include the application of catalytic filters~\cite{diesel_filter} for reduced exhaust fumes. Such solutions however, fail to address the non-exhaust emissions from diesel and non-diesel vehicles. The potential of road sweeping and washing to reduce non-exhaust related emissions was presented in a study in the Netherlands in 2010~\cite{netherlands_2010}. The authors, although they identify non-exhaust emissions as the main source for coarse PM in urban areas, conclude that their approach does not have a significant reduction in non-exhaust emissions. The benefits of ride-sharing to the environment have been discussed in various studies, such as~\cite{ride_sharing_2010, agent_based, beijing_sharing, carpooling}. However, these studies do not take a dedicated interest to non-exhaust emissions, but rather, to fuel consumption reduction. Fuel consumption reduction has been addressed with route suggestion solutions in~\cite{truck_route_2011, time_fuel_2013,fleet_fuel_2014}, for trucks and vehicle fleets. In~\cite{time_fuel_2013}, the authors present a linear programming solution to the Time-Dependent Pollution-Routing Problem. Fleets of vehicles are re-routed depending on traffic, and speeds are recommended based on emissions, driver costs, traffic and peak hour information. As a solution, the authors introduce a departure time and speed optimization algorithm. A similar approach for optimisation of fleet size is proposed in~\cite{fleet_fuel_2014}. In the same spirit, authors in~\cite{its_2013} study a variety of measures, such as traffic control, ban of heavy duty vehicles (HDV) and speed restriction, in order to achieve reduction of traffic related emissions. Traffic control (simulated simply by reducing traffic by 20\%) and HDV banning have a significant reduction in air-pollutants (20\% to 23\%), whereas speed control exhibits increase in PM emissions, due to HDV. Last but not least, use of electric vehicles, as an alternative to diesel and petrol ones, has been suggested for the reduction of traffic related air pollutants. In a feasibility study in Canada and Italy~\cite{team_play_2016}, the use of electric cars and electric motorcycles shows a reduction in CO$_2$ emissions, however, the study ignores non-exhaust related emissions, which are as relevant to electric vehicles as they are to diesel and petrol ones~\cite{non_exhaust_electric}. Another study in the city of Dublin ~\cite{dublin_ev_2011} uses home/work commute and traffic related data to study a number of electric vehicle market penetration scenarios and evaluates the emission decrease under each of them. However, only tailpipe emissions are taken into consideration, again overlooking brake and tyre wear and other non-exhaust emissions. As opposed to the majority of works that address the reduction of road traffic related emissions, we propose a traffic control and ride-sharing scheme, that reduces the amount of cars in the streets, and therefore the tyre-related emissions, as well as other non-exhaust and exhaust emissions. 

\section{Feedback-enabled Access Control}\label{sec:acs}

The pollution mitigation mechanisms discussed in the previous section, and the move from ICE to EV's that is so popular in many cities globally, is based on the assumption that the principal source of pollution is tailpipe in origin. As we have discussed in the previous section this assumption is at best only partially true and tyres, brakes, as well as road abrasion, may contribute significantly to PM generation. 
Additionally, EV's are generally heavier than ICE vehicles, potentially affecting tyre wear negatively. This means that the amount of emitted non-exhaust PM might actually even be elevated for EVs. Therefore, one must look for alternative mitigation mechanisms to combat these sources of PM generation. Apart from the obvious move from private to public transport or other modes of transport such a cycling and scooters, the only real viable mechanism is to develop an access control mechanism that is based on a feedback control strategy to regulate the safe levels of PM. It is one such strategy that we now develop.\newline 

\begin{figure}
\includegraphics[width=1\columnwidth]{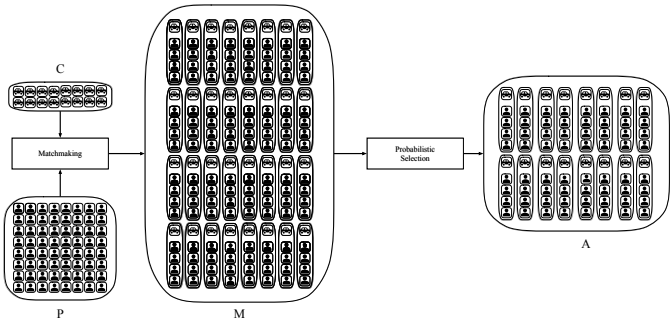}
\caption{Block model of the access control algorithm. First cars and passengers (respectively sets $C$ and $P$) get matched by a matchmaking algorithm. Once this process is over, each car and the corresponding passengers (set $M$), are randomly picked by a probabilistic method. The resulting set of cars, A represents the cars and passengers who are granted access to the city for the day.}
\label{fig:Scheme}
\end{figure}

Specifically, our objective is to maximise both the number of cars and people entering the city centre each day, while maintaining the tyre-generated PM emission levels significantly below the maximum permitted levels. The idea is to orchestrate an access control scheme so that it encourages ride-sharing. The access mechanism works in a simple way: at each day passengers are assigned to cars (drivers) through a matching method. Then, cars who want to have access to the city center are picked randomly using a probabilistic method that ensures fairness and privacy to each user and which is based on occupancy. See Figure~\ref{fig:Scheme}, for a visual depiction of this scheme.\newline

The rationale behind the choice of a probabilistic method instead of a deterministic one, like a water-filling algorithm, lies in the fact that the latter can be quite inefficient from the single user perspective. In order to use an access control scheme, an agent (driver or passenger) would typically buy a monthly or yearly access pass. This ticket provides them with the opportunity of competing with other users to access the city, either as a driver or as a passenger. Consider now the example of parents, that have to take their children to school outside the city centre before travelling into the city centre in the morning: even though they paid the same amount for a monthly or yearly parking ticket as everyone else, in a deterministic system they always  have a greater chance of missing out the chance of having access to the center of the city, as they arrive later than everyone else. Using a probabilistic system, as the one described in \cite{QoS}, we are able to guarantee equality in regards to access for all users over the long-term period of validity of their pass, irrespective of their constraints. \newline

For this Access Control Method, we assume that the controlled region (referred to as $R$) can accommodate up to $N$ vehicles per day, decided so that the tyre-related PM emissions are kept at low levels, ensuring thus that in general PM emission levels will remain low. There are mainly two challenges to make this method work efficiently.\newline

\begin{itemize}
\item[Q1 ] {\em Compliance:} How does one make sure that users comply with the matchmaking scheme, after access has been granted? \newline
\item[Q2 ] {\em Fair access:} How does one ensure that each driver is granted access to $R$ fairly with respect to other users (for instance, keeping the amount of average access the same among all cars)?\newline
\end{itemize}

We answer these questions in detail in the following subsections. 

\subsection{Ride-Sharing Compliance}

In a Ride-Sharing scheme, as the one described above, one of the crucial elements to make the architecture work is to ensure that both drivers and passengers comply with the matchmaking system. If users are not somehow punished for negative behaviour, they might be inclined to cheat the system to maximise their own personal advantage, which in turn might lead to sub-optimal results and to a poor Quality of Service (QoS) overall. As an example, in order to increase the probability of gaining access, a driver might accept as many drivers as possible and then refuse to pick them up; on the other hand, a passenger might choose to not show up, effectively wasting time and resources (the assigned seat). In this context, on the basis of the work done in \cite{DLTAndSocial} we propose the use of a digital token as a bond, or digital deposit, to ensure that passengers and drivers comply with their respective social contract (the matchmaking system). The risk of losing a token is then the mechanism that encourages agents to comply with these social contracts. There are multiple practical ways to implement this system: a possible example could be to have each user equipped with a digital wallet and the only way to participate to the matchmaking system is to have enough tokens to use as a bond. Another way could be to link the tokens to real money, so that losing a certain amount of them would result in a real economic loss for the agent. Note that the pricing of such tokens is beyond the scope of this paper and is dealt with in \cite{DLTAndSocial}. \newline

\begin{figure*}
\centering
\begin{tabular}{ccc}
\includegraphics[width=0.6\columnwidth]{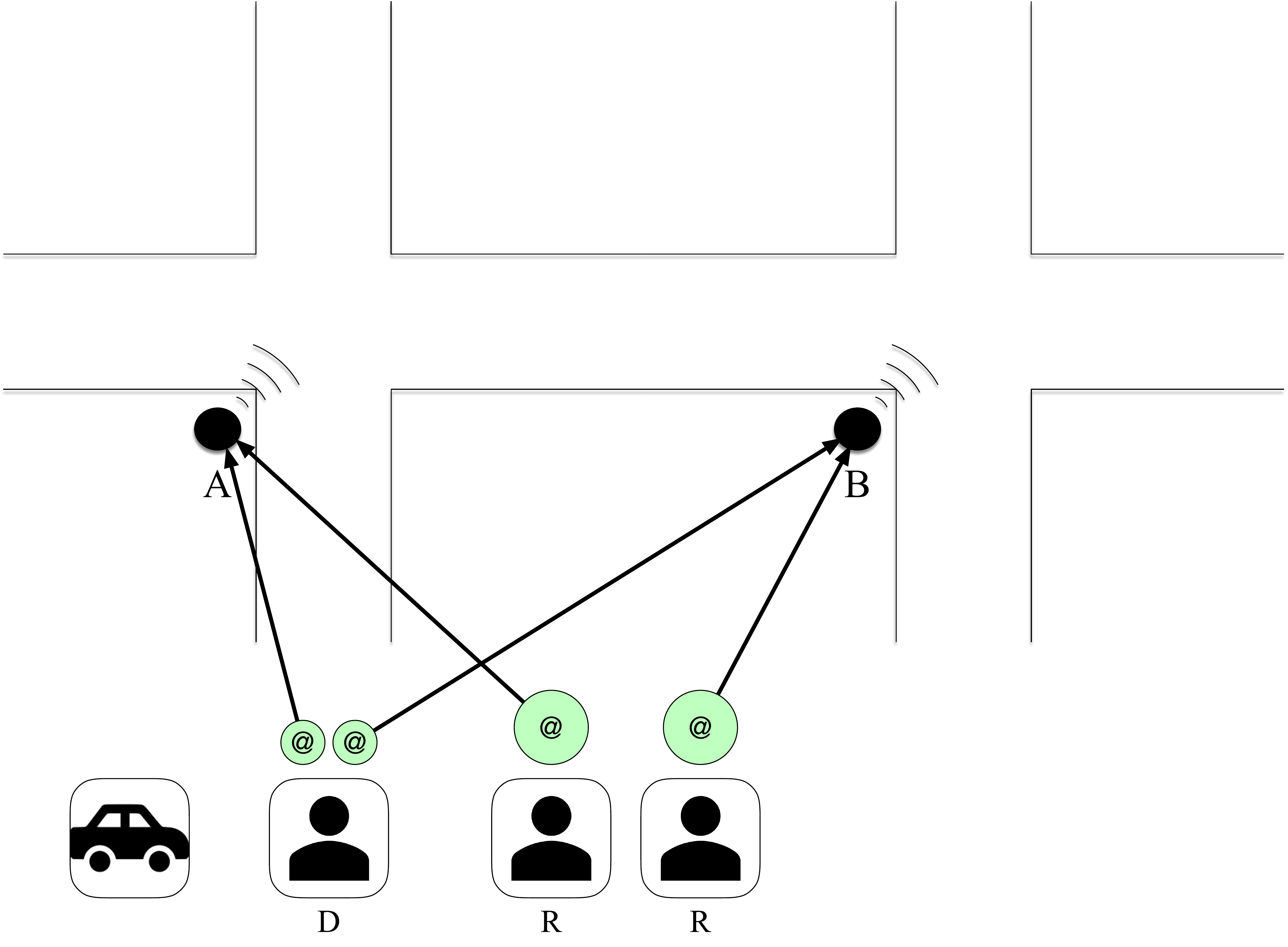} & \includegraphics[width=0.6\columnwidth]{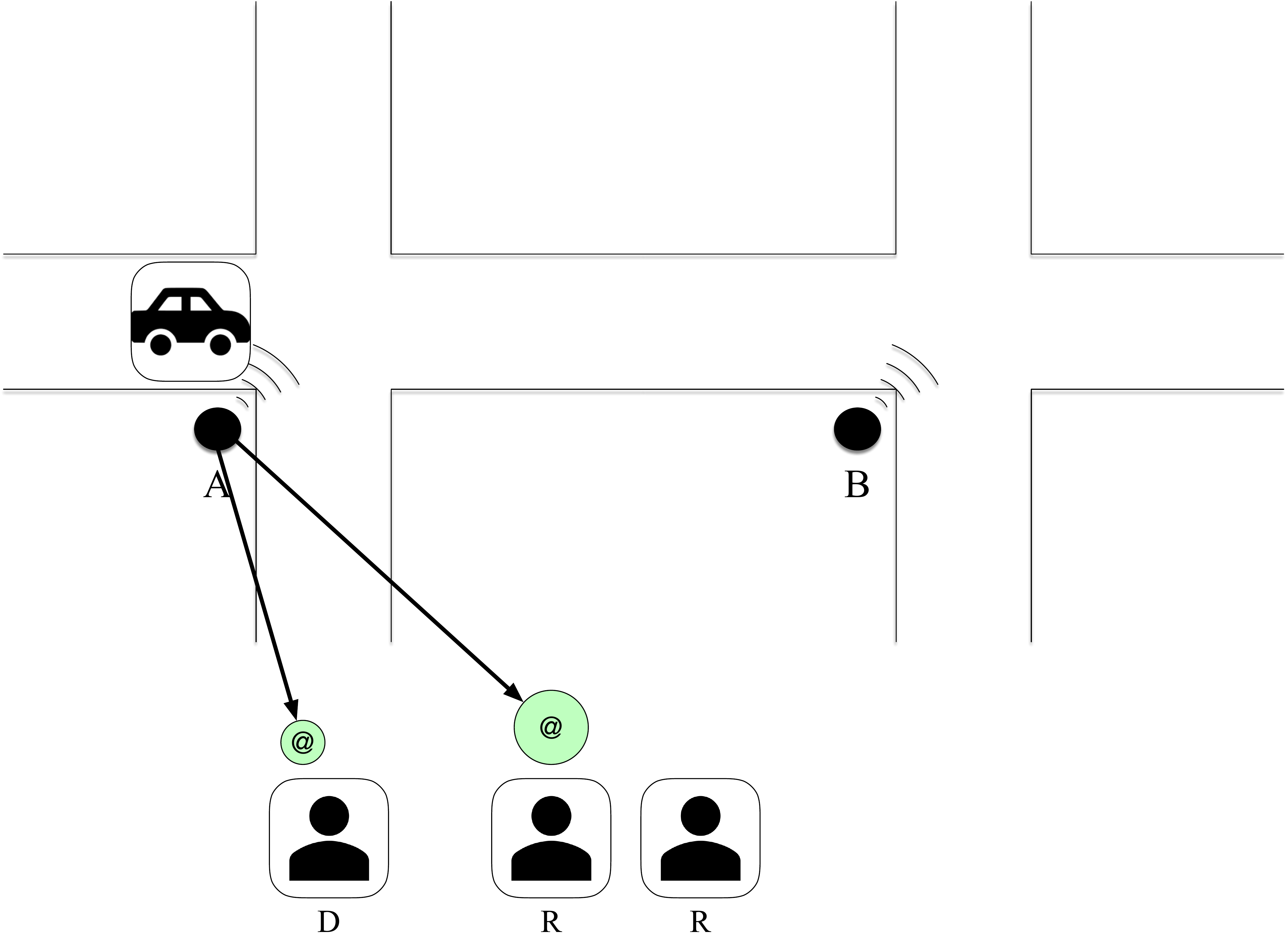} & \includegraphics[width=0.6\columnwidth]{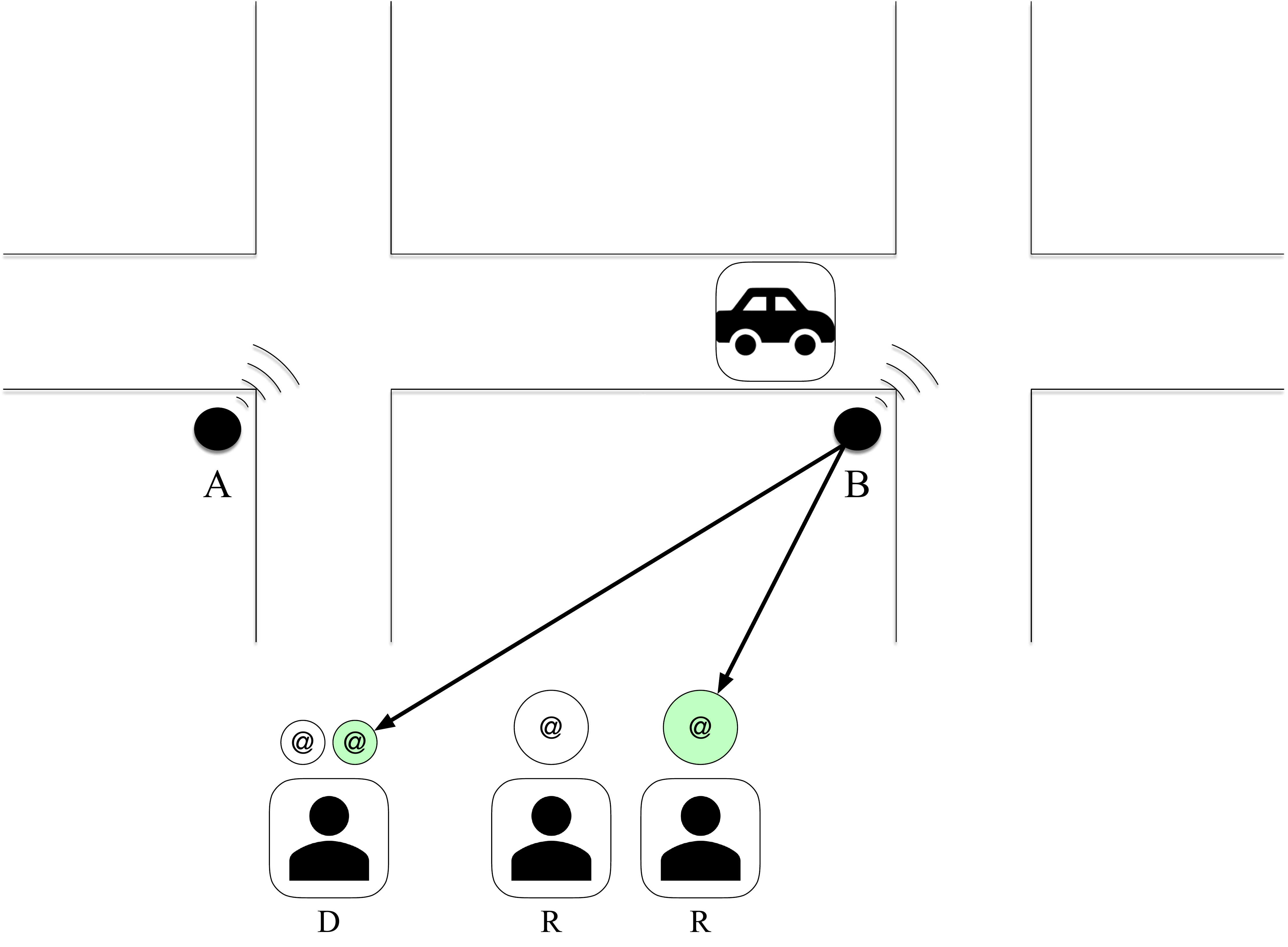}\\
(a) & (b) & (c)
\end{tabular}
\caption{(a) Driver $D$ deposits tokens for initiating a contract that they will pick up passengers from pick up points $A$ and $B$. At the same time, the two passengers deposit their tokens for appearing at the pick-up points. (b) Driver $D$ appears at pick-up point $A$ and collects the passenger from there, therefore, both the driver and the passenger retrieve their tokens for complying with the system. (c) Similarly, tokens are retrieved by the driver and the second passenger for both appearing at pick-up point $B$. At each stage the moving tokens are represented in green.}
\label{fig:compliance}
\end{figure*}

The simple idea is that, whenever a passenger is matched with a driver, they both agree on a specific \emph{pick up} point and on a time window. Once the passenger gains access to the city center, all the agents involved 'deposit' a \emph{token} to the designed pick up point (notice that this process is repeated between each driver and passenger, therefore a driver will deposit an amount of tokens equal to the number of passengers they are carrying). Then, in order to retrieve their token each agent needs to be physically present at the pick up point, in the designed time window. If unable to do so, the agent will forfeit the possession of the token that can be retrieved by any other passenger/driver present at that time and place. To have a better understanding of this process refer to Figure \ref{fig:compliance}. \newline
 
In what follows, we propose the use of a permissioned Distributed Ledger Technology (DLT) strategy to implement the proposed access control scheme. The acronym DLT is a term that describes blockchain and a suite of related technologies. From a broad perspective, a DLT is nothing more than a ledger held in multiple places, and a mechanism for agreeing on the contents of the ledger, namely the consensus mechanism. While this technology was first discussed in Nakamoto’s white paper in 2008 \cite{Nakamoto}, the technology has been used primarily as an immutable record keeping tool that enables financial transactions based on peer-to-peer trust \cite{Puthal}. In order to reach consensus, architectures such as blockchain operate a competitive mechanism enabled via mining (Proof-of-Work), whereas architectures such as the IOTA Tangle \cite{Zheng} based on Directed Acyclic Graph (DAG) structures often operate a cooperative consensus technique. The concept of using tokens to mark specific points where conditions are to be met, perfectly conforms with a DLT-based system. In fact, it is natural to use distributed ledger transactions to update the position of the tokens and to link them to the points of interest and associated data, using transactions (this can be done, for example, using smart sensors linked to digital wallets, as shown in Figure \ref{fig:compliance}). On top of that, a DLT-based system brings a number of advantages as a byproduct of its application to the smart city domain:\newline
\begin{itemize}
\item {\em Privacy :} In DLTs, transactions are pseudo- anonymous. This is due to the cryptographic nature of the private address\protect\footnote{https://laurencetennant.com/papers/anonymity-iota.pdf}, which is less revealing than other forms of digital payments that are uniquely associated with an individual \cite{cryptoprivacy}. This does not mean that DLTs users' identities are completely anonymous, especially in architectures in which it is possible to follow the trail of transactions among addresses. At the same time though, DLT systems are pseudo-anonymous in the sense that they manage to hide the details of single users and through randomization of the address they can make it difficult for attackers to trace the transactions. Therefore, from a privacy perspective, the use of DLT is desirable in a smart mobility scenario.\newline

\item {\em Ownership :} Transactions in the DLT can be encrypted, thus allowing every issuer to maintain ownership of their own data. In the aforementioned setting, the only information required to remain public is the current ownership of the tokens, whereas auxiliary information (e.g., user quality of service, statistics on the usage of the system) can be encrypted. This information can later be monetized for the benefit for the data owner.\newline

\item {\em Microtransactions :} Due to the amount of vehicles in an urban environment, and due to the need of linking the information to real time conditions (such as traffic or pollution levels), there is the demand for a fast and large data throughput.
\end{itemize}

Furthermore, the DLT system needs to be designed in a way such that whenever a user issues a token as a bond, that same user can retrieve the token \emph{if and only if} they are present at the pick up zone at the designed time. To do so we make use of the same mechanism and architecture proposed in \cite{NarendraYale}: namely a Proof of Position (PoP),  DAG-based DLT called \emph{Spatial Positioning Token} (SPToken). Unlike other DLTs, in which each user has complete freedom on how to update the ledger with transactions, the SPToken network has a regulatory policy based on the physical positions of agents. This feature allows for a number of different uses: it can be employed  to prevent agents to add transactions that do not possess any relevant data (since transactions can be encrypted)\cite{NarendraYale} or, as in this specific paper, it can be used to make sure that an agent satisfies certain conditions. Therefore, as a validation mechanism, SPToken makes use of PoP to authenticate transactions. In other words, for a transaction to be authenticated, it has to carry proof that the agent was indeed at the pick up point, at the designated time. This is achieved via special nodes called \emph{Observers} (see Fig.~\ref{fig:compliance}). Each observer is linked to a physical sensor in a city and it acts as a witness for the transaction. A sensor can be a fixed piece of infrastructure, or a trusted vehicle whose position is verified. As soon as a car is granted access to $R$, each user will deposit their tokens at the designated pick up zone. As soon as an agent reaches in time their pick up point, where one or more of her tokens are available to be picked, a short range connection is established (e.g., via Bluetooth) with the observer (whose job is to authenticate the transaction) and the token is transferred back to the owner's account. Refer again to Figure \ref{fig:compliance} for a better understanding of this process. This mechanism ensures that users have to be physically present in the interested locations to be able to retrieve their bond. This further authentication step makes SPToken a permissioned DAG-based DLT (similar to permissioned blockchains \cite{Puthal}), i.e., a distributed ledger where a certain amount of trusted nodes (the observers, in this case) is responsible to maintain the consistency of the ledger (as opposed to a public one, where security is handled by a cooperative consensus mechanism \cite{DLTAndSocial}).\newline

{\bf Comment :} Before continuing, we want to stress that very often in the context of \emph{smart cities}, algorithms assume full compliance with policies that are designed to optimise the resource allocation. To assume that a human agent would not break rules, especially if an individual profit can be made, is a very strong hypothesis that if relaxed might lead the whole system to fail and to produce less than optimal results. Therefore, it is the authors' opinion that the use of a compliance system is of paramount importance in the setting described so far, if efficiency is to be achieved. The issue of compliance is often overlooked.

\subsection{Mechanism Description}

We consider now the problem of allocating a certain amount of resources (i.e., permitted number of cars) among a set of agents (i.e., drivers and passengers using the scheme).  The proposed method is inspired by the algorithm presented in \cite{QoS}, appropriately adjusted to the requirements of our ride-sharing scheme.\newline

We consider the following scenario. There is a population of size $n$ of citizens participating in the scheme, who request to commute to $R$ on a daily basis. The controlled region can accommodate up to $N$ vehicles per day. We assume $n > N$ and the population could be either passengers or drivers. We assume that there is a fleet of $N' > N$ electric vehicles in the scheme that are requested by the population for access in $R$, with $n > N'$. Without loss of generality and to facilitate presentation of our mechanism, the entities \textit{driver} and \textit{car} are considered equivalent and the corresponding terms are thus used interchangeably.  As already mentioned in a previous section, our method is organized in two phases : matchmaking and probabilistic access. During matchmaking, we match passengers with drivers and group them into cars. The matching can happen in a number of ways, depending on the specific requirements of those who apply the system. For example, passengers could be matched with drivers based on proximity of their departing/arriving area, or based on a preference priority ranking that drivers/passengers maintain for each other. In our simulations we take a simple approach and match passengers randomly with drivers (and subsequently with cars), as long as there are available seats in the vehicles,  taking into consideration the frequency at which a particular passenger has been assigned a seat in the past. That is, if a passenger has been assigned a seat less than $50\%$ of the time, then they are given priority to take a seat in a car, otherwise, they are not given priority. After the matchmaking is complete, each car is assigned an access probability based on its occupancy records. All cars with high enough probability, are permitted access to the city center. We present the technical details of this procedure, next.\newline

In our system, we will use $k$ to denote number of days (i.e., $ k=0,1,2,3,\dots$). For ease of interpretation we assume that access is granted on a daily basis to each user, but the algorithm is not affected by this assumption. Then, $X_i(k)$ is the state variable associated with each driver; it takes the value $1$ if the $i$th driver is given access to $R$ on the $k$th day and zero otherwise. Thus, $\overline{X}_i(k)$ is the average access for the $i$th driver up to the $k$th day, defined as
\begin{equation*}
\overline{X}_i(k) = \frac{1}{k+1} \sum_{j=0}^{k} X_i(j).
\end{equation*}
In the above context, let $z_i \in [0,1]$ represent the frequency of accessing the city for a car $i$, and $f_i : [0,1] \rightarrow \R$ be a convex cost function associated with it, representing the car's priority during the second phase of our mechanism. In this context the shape of this function can take into account a variety of factors: the amount of money paid for the pass (e.g., premium and standard account), the amount of public transportation available in the area where this user lives or the type of vehicle driven. Following~\cite{QoS}, we are interested in solving the following shared-resource optimization problem,
\begin{eqnarray}
\label{eq:optprob}
\minimize_{z_1,\ldots, z_{N'} \in \R} \,\, &\sum_{i=1}^{N'} f_i (z_i) \nonumber\\
\text{subject to} \,\, & \sum_{i=1}^{N'} z_i = N,\\
& z_i\geq 0, \quad i=1,\ldots, N'.\nonumber
\end{eqnarray}
Our aim is then to control the value of the variable $X_i(k)$ (i.e., the access to $R$, at each time step) in such a way that the average access of user $i$, $\overline{X}_i(k)$, converges to the optimal value $z_i^*$, subject to  $\sum_{i=1}^n X_i \approx N$ (notice that we are not requesting the algorithm to exactly match the required amount of cars, at each time step but we are instead interested in obtaining $\text{lim}_{k,\infty}\sum_{i=1}^n \overline{X}_i(k) = N$).  In order to do so, the probability that at each time step car $i$ gains access to the city center (i.e., $X_i(k) = 1$) is ruled by the following equations:
\begin{equation}
 \label{eq:pi}
p_i(k) \triangleq \P(X_i(k) = 1) = \Gamma(k)\ \frac{\overline{X}_i(k)}{f_i'(\overline{X}_i(k))}\dfrac{n_i(k)}{c_i},
\end{equation}
\begin{equation}
    \label{eq:alg3}
    \Gamma(k+1) = \Gamma(k) + \alpha \left( N - \sum_{i=1}^{N'} X_i(k) \right),
\end{equation}
where $n_i(k)$ is the number of passengers carried in car $i$ at time $k$, $c_i$ is the car's maximum capacity and $\Gamma(k)$ is a global scaling variable, dependent on the parameter $\alpha >0$, whose dynamics ensures $p_i(k) \in [0,1], \forall i,k$.  Notice that, equation (\ref{eq:pi}) differs from the one proposed in \cite{QoS} by the factor $n_i(k)/c_i$: since we are interested in maximising the amount of people getting into $R$ (while maintaining the amount of users having access close to $N$), this factor ensures that a fully filled car will have higher probability to be granted access than an empty one. As a further element, notice that in a classical setting, the presence of $\Gamma(\cdot)$ requires the existence of a centralised entity to compute and broadcast this global variable to all the agents in the network. In a  DLT-based system, on the other hand, where informations are stored in a public ledger, the value $\Gamma(\cdot)$ can be computed independently by each user, therefore the algorithm can be executed in a completely decentralised fashion. A discussion on the convergence of this algorithm is beyond the scope of this paper and the interested reader can refer to \cite{QoS} for further details.

\section{Simulations and Results}\label{sec:simulations}
We now present empirical results to illustrate the efficacy of the techniques presented in the previous section. In what follows we based our simulations on the recent report~\cite{canal_dublin}. We assume a city of the size of Dublin in Ireland, with population $1,100,000$ approximately, of which $50,000$ are considered drivers and about $400,000$ are daily commuting passengers\footnote{In the report~\cite{canal_dublin} it is stated that between 7-10 am, about 210,000 commuters entered the city center. We make the assumption that in the length of the day that number can potentially double and therefore consider a population of $400,000$ commuters}. Consequently, we have a fleet of $50,000$ EV's, out of which only $40,000$ are permitted in the city centre $R$ on a daily basis\footnote{In the report~\cite{canal_dublin} it is stated that between 7-10am, about 50,000 cars entered the city center. Therefore, we limit the number of drivers to that number and the number of permitted cars to slightly less than this figure.}. All users that are not granted access to the city on a $EV$, are redirected to use public transportation. In our simulations, we set $\Gamma[0] = 1$, that is, the value that the parameter $\Gamma$ takes the first day of the scheme's operation, and $\alpha = 0.0001$. We also consider an application period of 360 days, that is slightly less than a year long. For convenience, on the first day of the operation, we consider that all drivers are permitted access.\newline 

\begin{figure}
\begin{tabular}{c}
\includegraphics[scale=0.18]{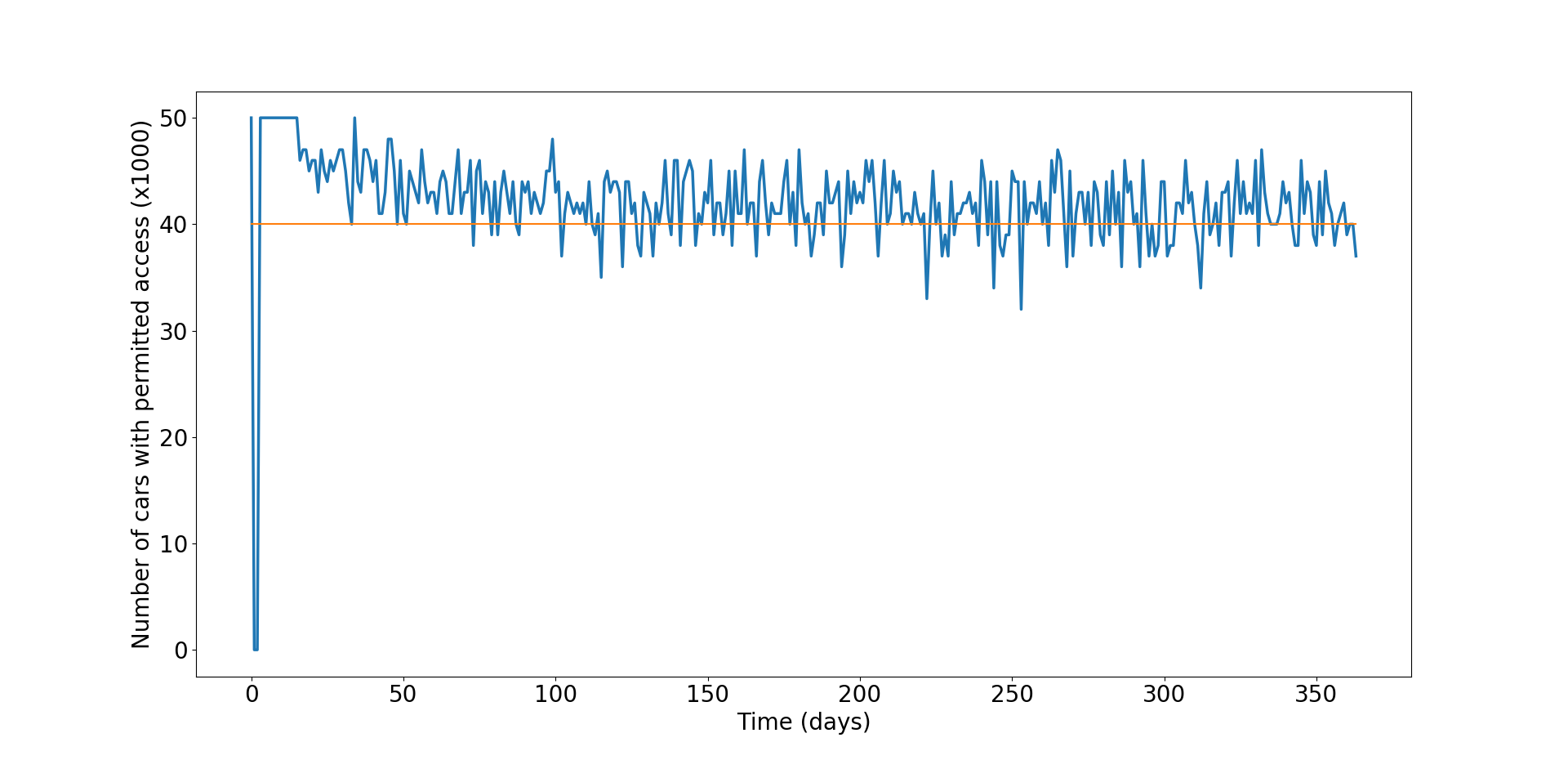} \\(a) \\ \includegraphics[scale=0.18]{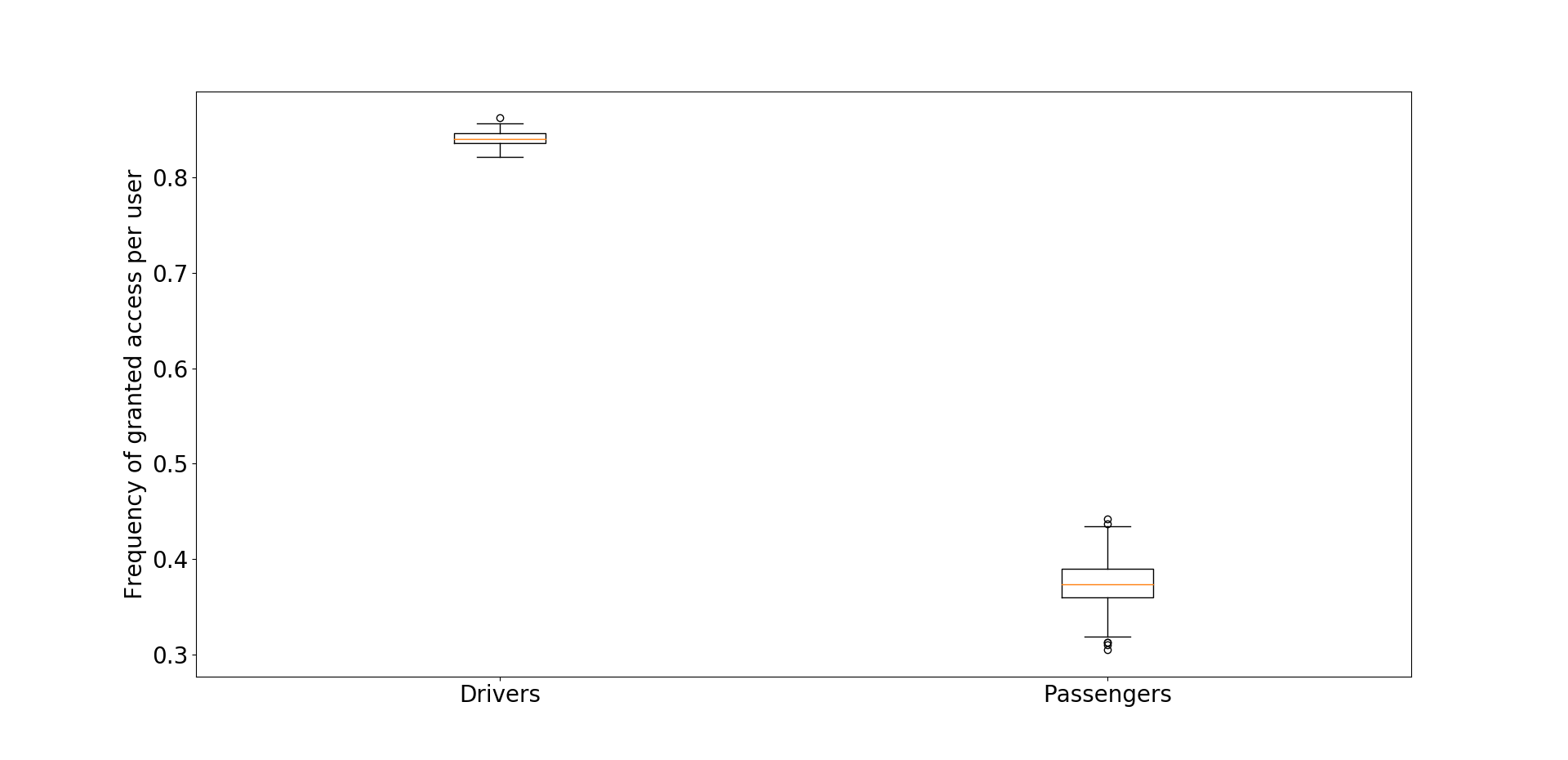}\\
(b)
\end{tabular}
\caption{(a) Number of cars with granted access in the length of a year (b) Frequency of granted access per user in the scheme}\label{fig:simulation_city}
\end{figure}

The simulation results are presented in Figures~\ref{fig:simulation_city}(a) and (b). Fig.~\ref{fig:simulation_city}(a) shows the number of cars that are granted access every day. Although at the beginning the number of cars in area $R$ are above the maximum permitted number ($40,000$), due to the effect of the access control mechanism this value is quickly reduced, stabilizing around the maximum level, on average, for the rest of the application period. Notice that if the maximum number of drivers in the city center was a hard constraint, it would be sufficient to reduce $N$ to take into account the fluctuations around this value. In Fig.~\ref{fig:simulation_city}(b), we show the frequency of being granted access, per user, on average over a period of one year. The small variance indicates that each user is granted fair access to the system. Regarding the commuting (shared) passengers, every passengers is granted access more than 1/3 of the time.\newline 

\begin{figure*}
\begin{center}
\begin{tabular}{ccc}
\includegraphics[width = .7\columnwidth]{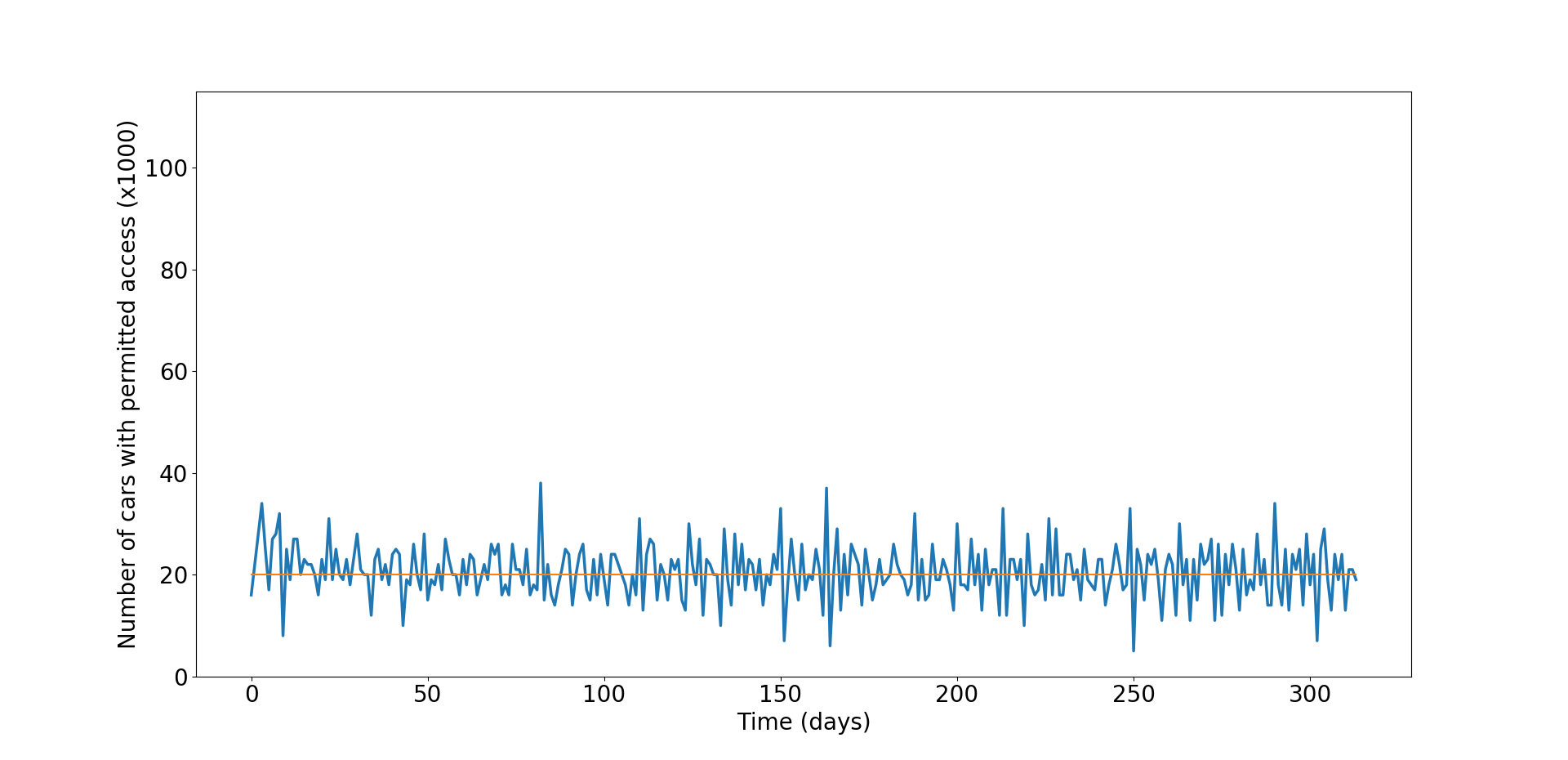} & \hspace*{-1cm}\includegraphics[width = .7\columnwidth]{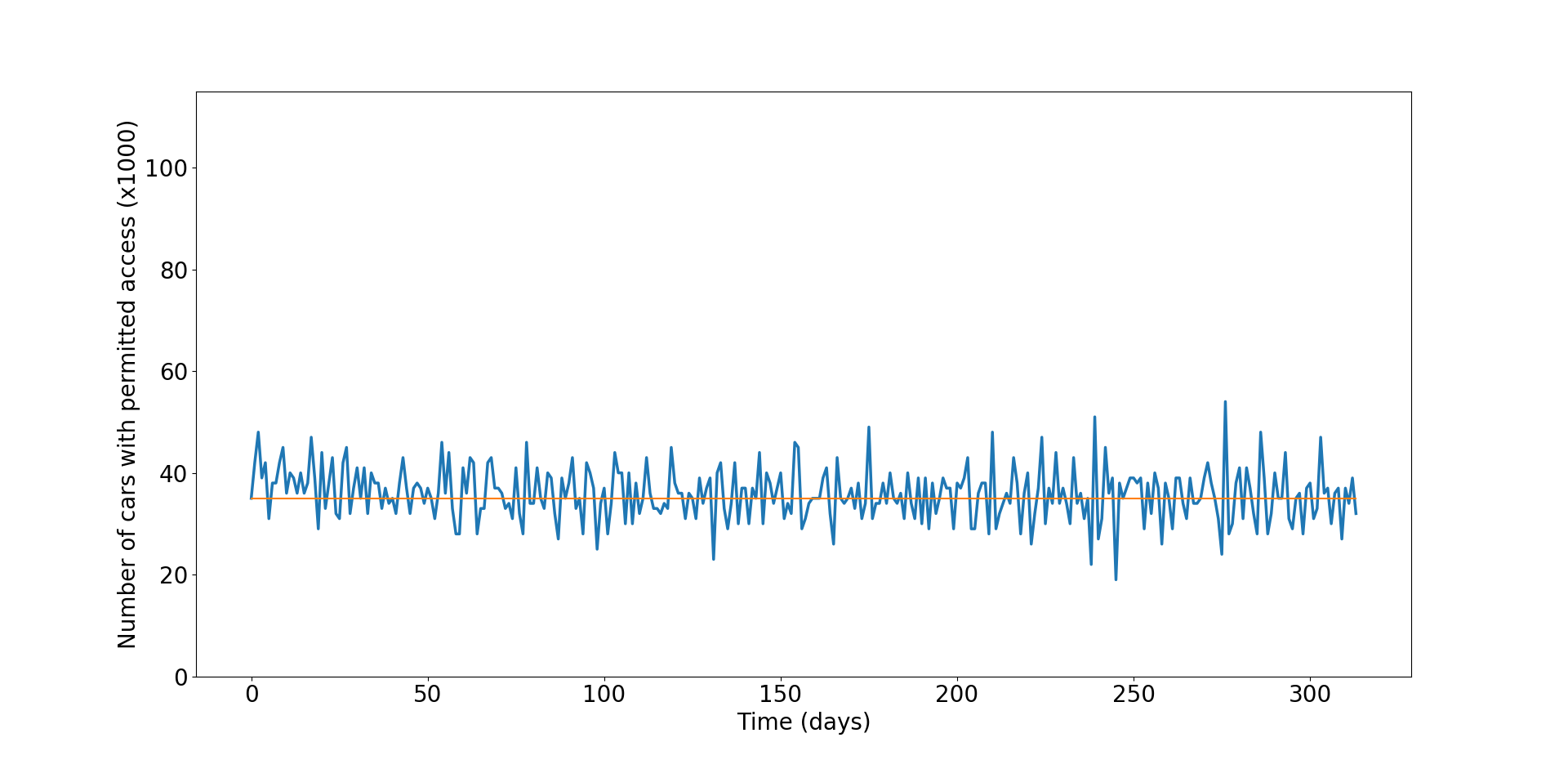} & \hspace*{-1cm}\includegraphics[width = .7\columnwidth]{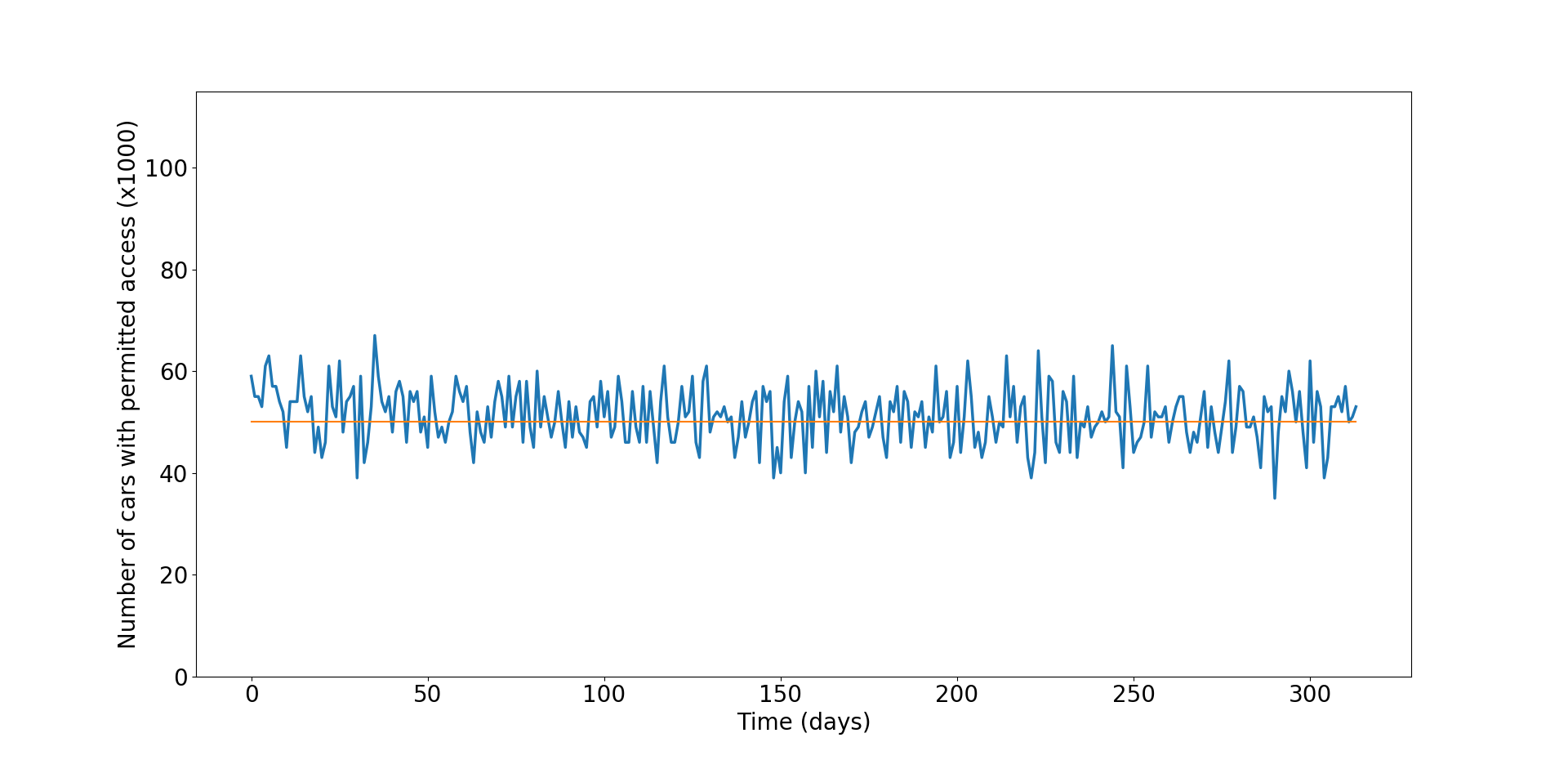}\\
(a) Max = 20k cars & \hspace*{-1cm}(b) Max = 35k cars & \hspace*{-1cm}(c) Max = 50k cars\\ \includegraphics[width = .7\columnwidth]{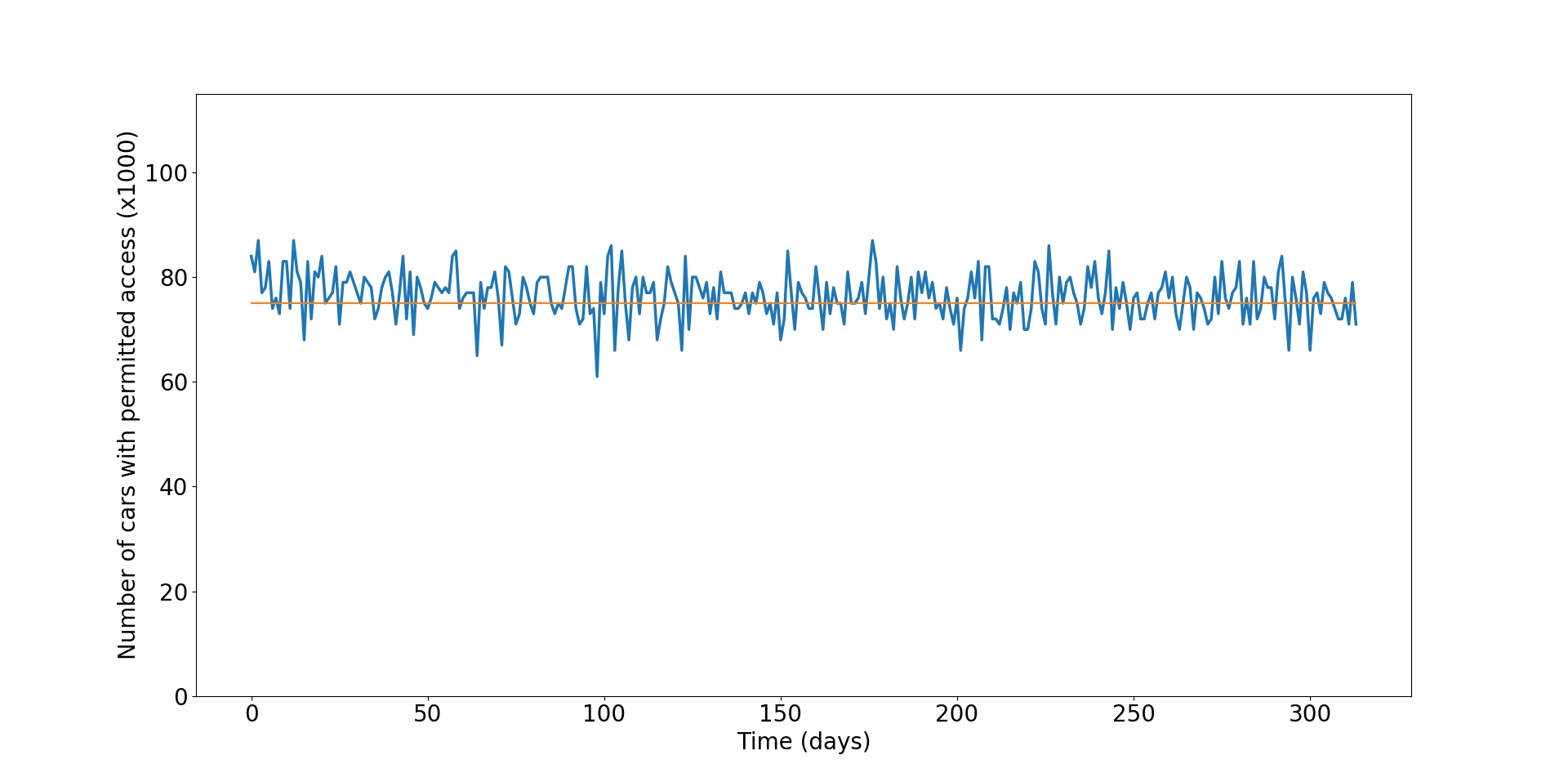} & \hspace*{-1cm}\includegraphics[width = .7\columnwidth]{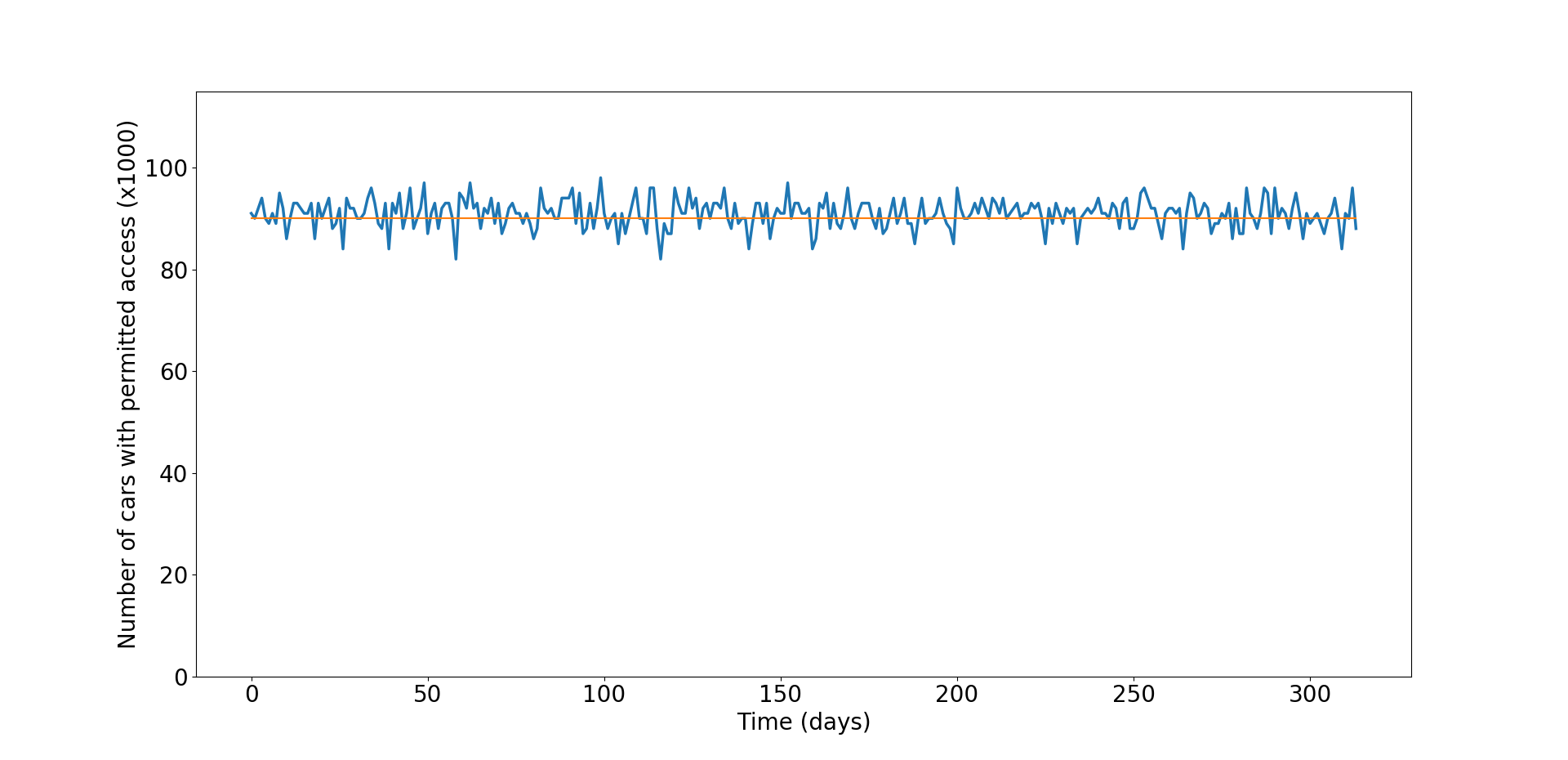} & \hspace*{-1cm}\includegraphics[width = .7\columnwidth]{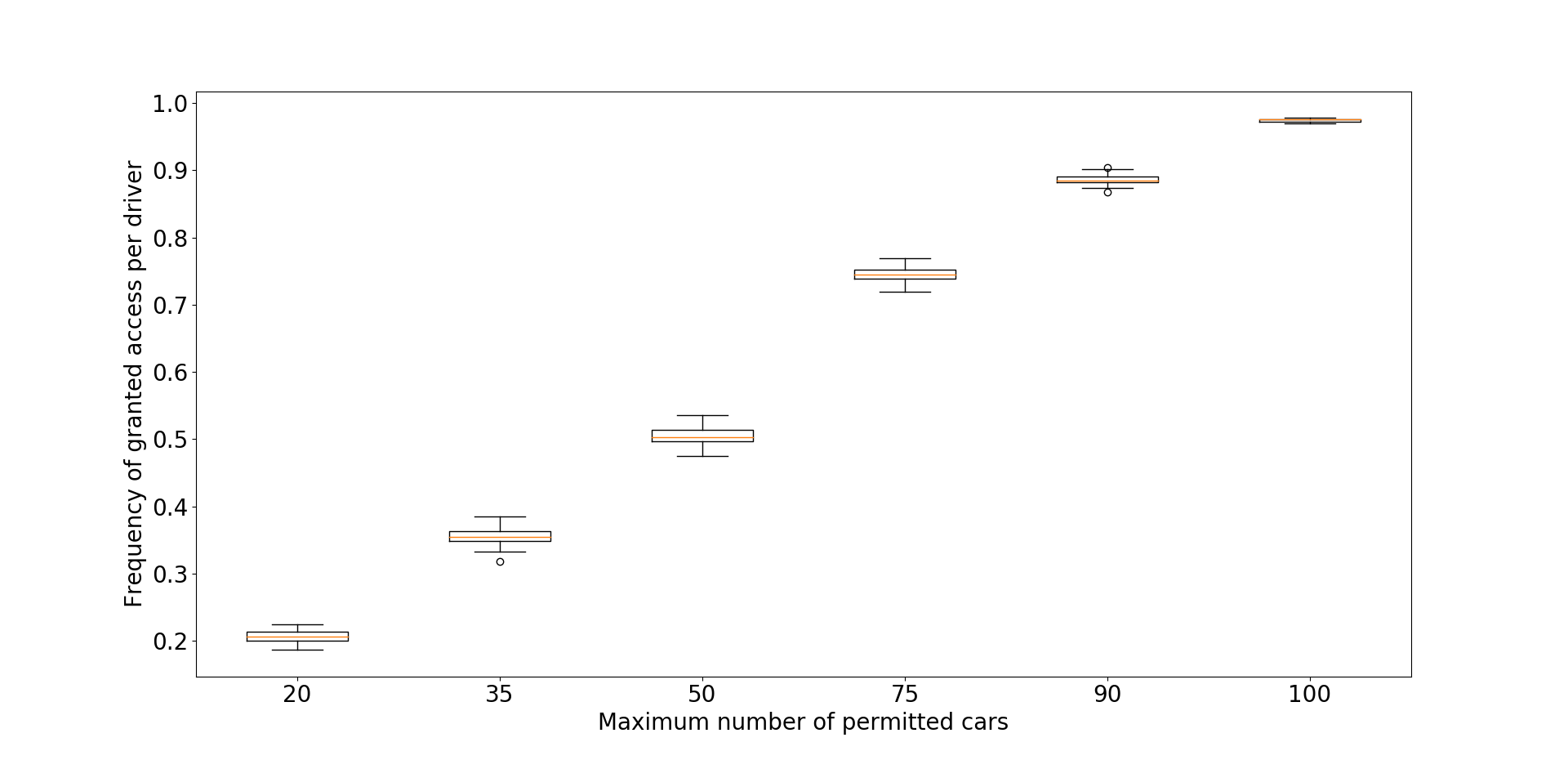}\\
(d) Max = 75k cars & \hspace*{-1cm}(e) Max = 90k cars & \hspace*{-1cm}(f)
\end{tabular}
\caption{(a)-(e) Amount of permitted cars, for varied values of maximum allowed vehicles. Steady state values depicted. (f) Frequency of granted access over a year, per driver, for different setting of number of allowed vehicles.}\label{fig:varied_access}
\end{center}
\end{figure*}

Figures~\ref{fig:varied_access}(a)-(e) depict the number of cars with access, when the number of maximum permitted drivers changes and all other parameters in the system remain the same. The plots depict the steady state values. We observe that in all cases, the number of cars with granted access converges to the maximum value, on average. In terms of fair access, we show in Fig.~\ref{fig:varied_access}(f) boxplots of the frequency at which \textit{each driver} is granted access to the scheme, in the length of a year, with regards to the maximum number of cars permitted in $R$. As expected, the frequency increases as the available amount of resources increases. We highlight that in all cases, the variance is very small, meaning that all drivers in the scheme are ensured fair access (i.e., all drivers are able, on average, to access the city center the same number of times).\newline 


\begin{figure}
\begin{center}
\includegraphics[scale=.18]{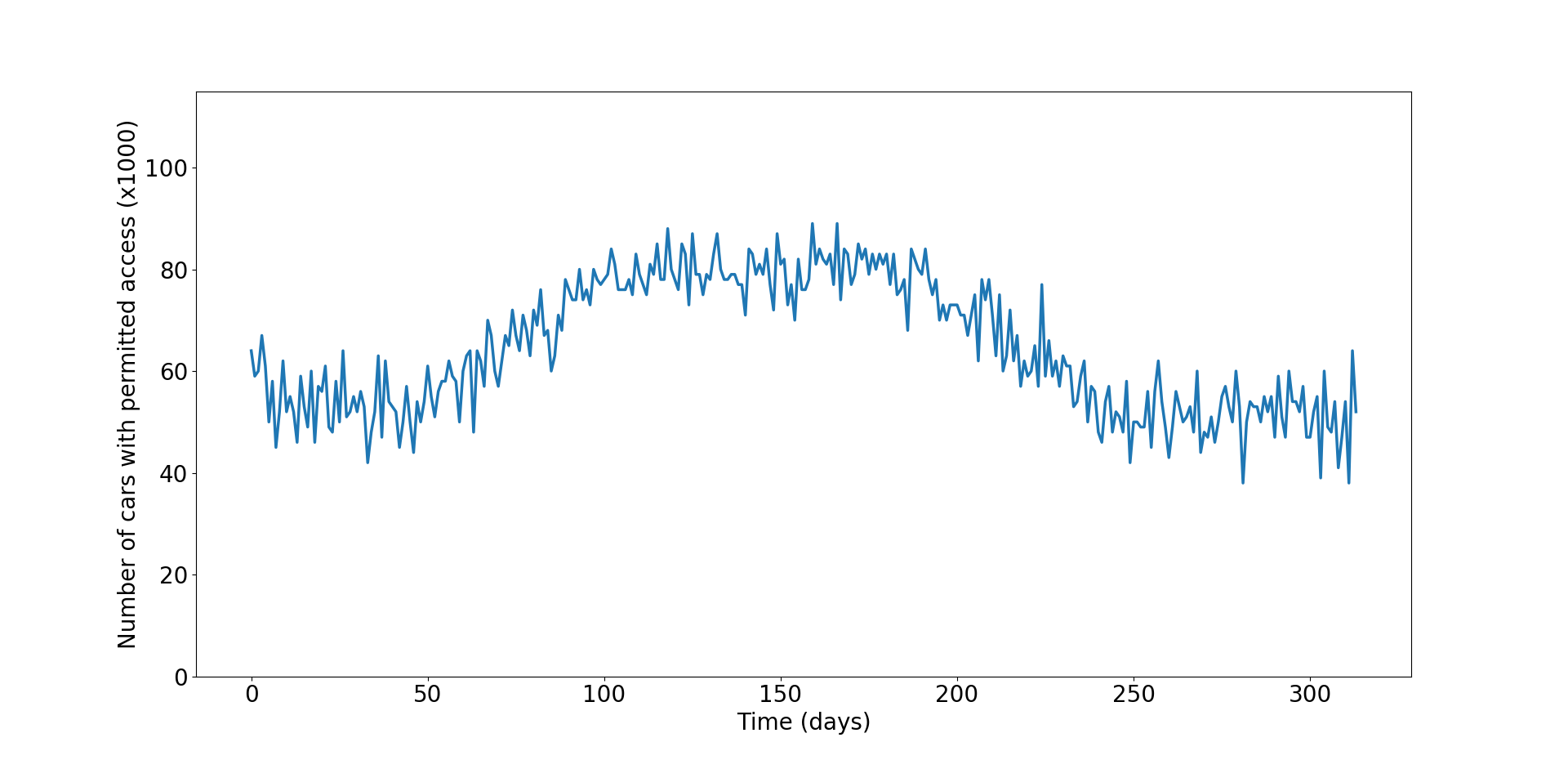}
\caption{Amount of permitted cars, while changing the value of maximum allowed vehicles gradually from $N = 50k$ to $N = 80k$ and vice versa.}\label{fig:changing_N}
\end{center}
\end{figure}

Finally to prove the efficacy of our approach in a more dynamic setting, we allow the number of maximum permitted cars to vary during the year. There are many reasons that make this a realistic scenario: the city municipality might wish to increase the number of permitted vehicles for the holiday seasons, or reduce it during heat waves, for example. We simulate this setting by changing the number of allowed cars during the year and we present the results in Figure~\ref{fig:changing_N}. Here, for the first two months of the operating period (i.e., 60 days) we give access to $N = 50k$ vehicles. For the next forty days $N$ increases linearly and for in the interval (of days) $[100, 180]$ it is set at $N = 80k$ vehicles. After that, the number of permitted cars decreases linearly again until it is set to the initial value, $N = 50k$, for the rest of the operating period. As we observe in the plot, our system reliably controls the access of vehicles, maintaining the number of permitted cars on average at a stable level around the set of maximum values.\newline

\begin{figure}
\centering
\begin{tabular}{c}
\includegraphics[scale=0.4]{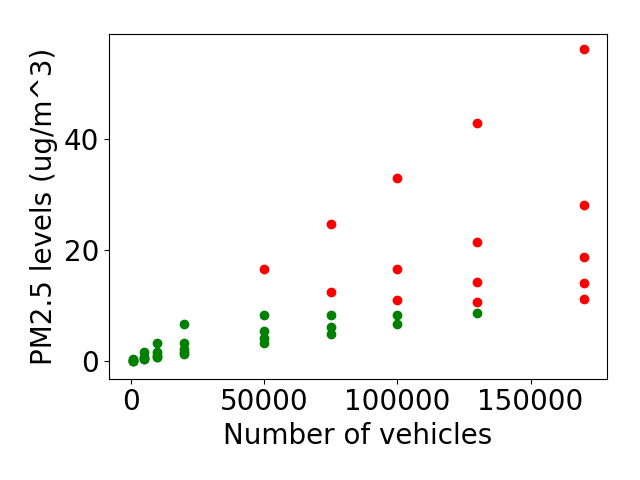} \\ (a) \\ \includegraphics[scale=0.4]{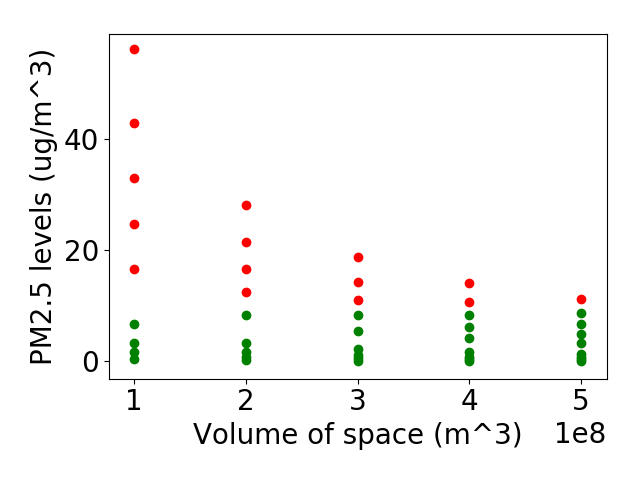}\\
(b)
\end{tabular}
\caption{Levels of tyre-wear related $PM_{2.5}$ emissions per number of vehicles operating (a) and volume of space where the matter is dispersed (b)}\label{fig:pm_cars_space}
\end{figure}

Regarding pollution levels caused by the $PM_{2.5}$ pollutant coming \textit{just} from the tyre wear of vehicles, we present in Figures~\ref{fig:pm_cars_space} the amount of particulate matter, depending on two variables: number of cars permitted in a city (Fig.~\ref{fig:pm_cars_space}(a)) and the volume of road network in a city (Fig.~\ref{fig:pm_cars_space}(b)). With regards to the volume of a city's road network, we wish to estimate, very approximately, the air-space in which the airborne PM is dispersed. For this, we assume that the total mass of PM generated per hour becomes uniformly dispersed throughout a volumetric space which is determined by the street length, an average street width of 10m and effectively enclosed by an average building height of 4m. Furthermore, we assume rather simplistically that the air in this volume is continuously replenished with an equivalent volume of fresh clean air at a rate of one air change per hour, in such a way as to maintain a pollution level which remains effectively constant with time. For the length of the road network, we can compute the total length of the streets in a predefined area in a city. Note that even though $4m$ is a somewhat arbitrary number for these simulations, the basic points remain valid irrespective of this assumption; that the amount of tyre generated PM can be regulated using our access control method. To this end, and based on the above assumptions,  Fig.~\ref{fig:pm_cars_space} depicts the amount of PM per $m^3$ as a function of the number of vehicles operating in a city, per possible volume of space (computed as described above). In these figures, we depict in green the levels deemed safe for human health (i.e., the ones below the maximum permitted levels) and in red the ones exceeding the annual permitted levels. These plots suggest that, with the present situation in Dublin city (that is, ~500,000 cars out of which ~170,000 change tyres every year, and a space volume of approximately $450,000,000m^3$), the levels of tyre-wear related $PM_{2.5}$ emissions are very high. However, applying an access control scheme that restricts the number of vehicles to \textit{at most} 100,000 vehicles per day, can maintain the PM levels at acceptable levels even in small size cities with relatively small volume of space.

\section{Conclusions}\label{sec:conclusion}

The contributions of this paper are divided into two sections. In the first one, a detailed data analysis shows suggests that a simple ban on ICE vehicles does not address the problem of non-exhaust emissions (PM from tyres, in particular). Although there have been previous studies that present such numbers for other cities, we emphasise the point that in Dublin, the PM levels from tyres alone might be above the levels that are deemed safe by WHO. This provides us with the rationale to introduce, in the second part, an access control and ride-sharing scheme to limit the amount of cars in cities and therefore maintain the amount of airborne PM  within safe levels for our health. This system is designed in such a way to encourage users to comply with the matchmaking scheme and to guarantee fair access to each car. Finally, to validate the proposed algorithm, we make use of extensive simulations to show that each user receives fair access to the city centre and that the PM emissions are kept within safe boundaries. As for future lines of research we will further extend the present work by using more complex models for tyre abrasion and airborne diffusion to obtain more accurate estimates for non exhaust emissions.

\begin{wrapfigure}[7]{l}{.3\columnwidth}
\centering
\vspace*{-.3cm}
\includegraphics[width = .3\columnwidth]{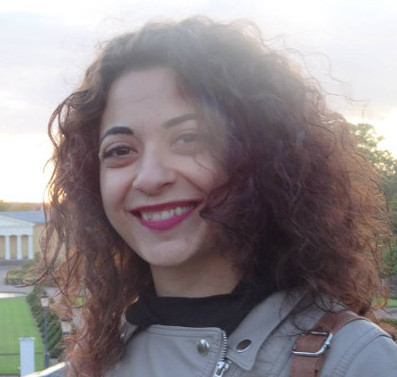}
\end{wrapfigure}
\textbf{Panagiota (Yota) Katsikouli} received the Diploma and MSc in Computer Engineering and Informatics from the Polytechnic University of Patras, Greece, in 2011 and 2013 respectively, and the PhD in Informatics from the University of Edinburgh, Scotland, in 2017. She is currently a post-doctoral researcher with the University College of Dublin.
Her research interests include human mobility, smart mobility, analytics for mobile data, distributed algorithms for mobility data.

\begin{wrapfigure}[7]{l}{.3\columnwidth}
\centering
\vspace*{-.3cm}
\includegraphics[width = .3\columnwidth]{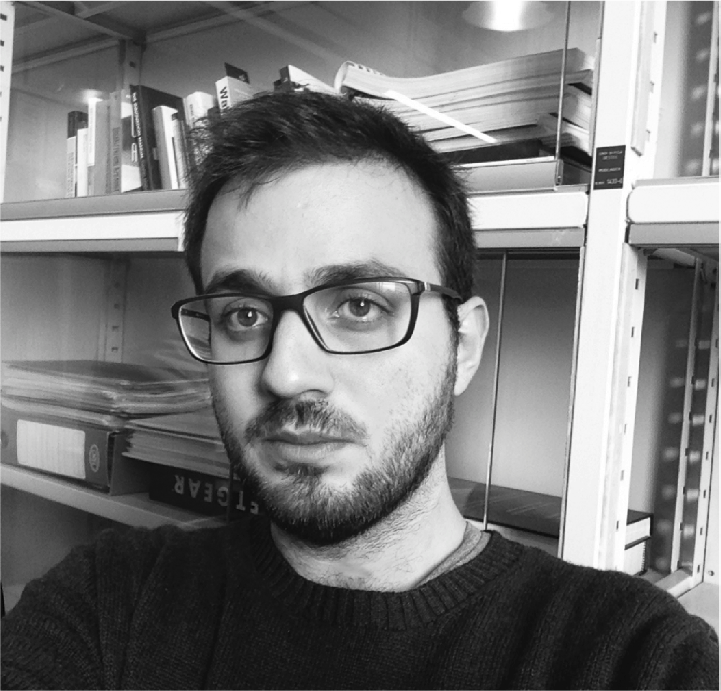}
\end{wrapfigure}
\textbf{Pietro Ferraro} received a PhD in control and electrical engineering from the University of Pisa, Italy, in 2018. He is currently a Post Doc Fellow with the school of electrical and electronic engineering at University College Dublin (UCD). His research interests include control theory applied to the sharing economy domain.

\begin{wrapfigure}[7]{l}{.3\columnwidth}
\centering
\vspace*{-.3cm}
\includegraphics[width = .3\columnwidth]{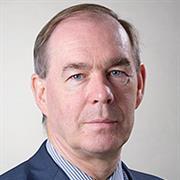}
\end{wrapfigure}
\textbf{David Timoney} received the Ph.D. degree in combustion modeling in diesel engines from UCD where he is currently pursuing the degree in mechanical engineering. He was with Ricardo Consulting Engineers plc., U.K. He joined UCD as an Assistant Professor in 1981, where he has been involving in thermodynamics and energy conversion systems. His research activities have been focused on internal
combustion engines and on transport-related energy topics. He is a member of the Society of Automotive Engineers, a Chartered Engineer and a fellow of the Institution of Engineers of Ireland, and a fellow of the Irish Academy of Engineering.

\begin{wrapfigure}[7]{l}{.3\columnwidth}
\centering
\vspace*{-.3cm}
\includegraphics[width = .3\columnwidth]{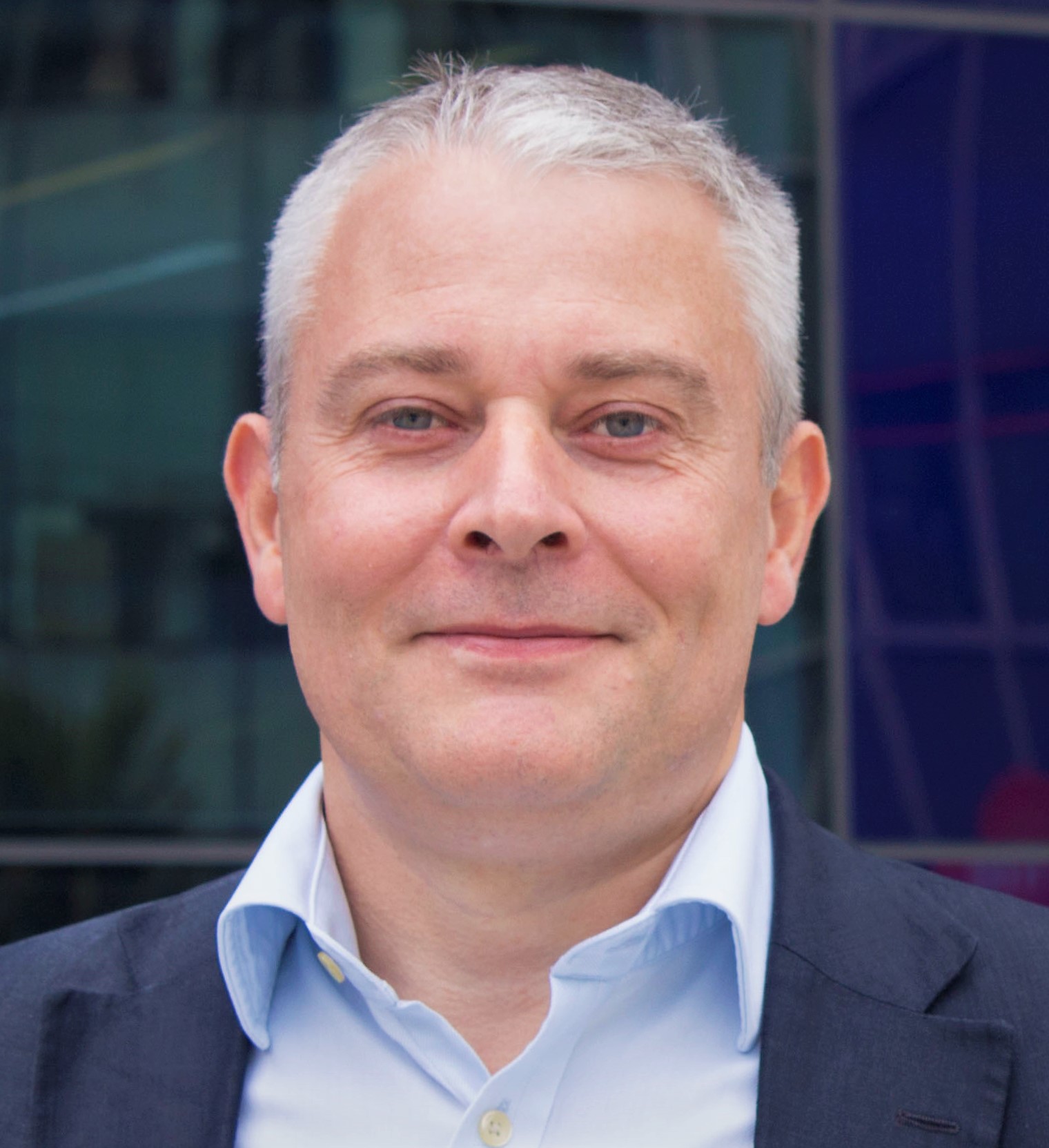}
\end{wrapfigure}
\textbf{Marc Masen} is an Associate Professor in Tribology at Imperial College London and the educational lead in Mechanical Engineering Design. He holds a PhD from the University of Twente, the Netherlands, where he is also a visiting researcher. He is a past Chair of the Institute of Physics Tribology Group and the conference co-chair of the biannual ICoBT conference-series on BioTribology. Marc serves on the editorial boards of the Proceedings of the Institution of Mechanical Engineers Part J: Journal of Engineering Tribology, and the journal Biotribology. His research interests include the friction and wear behaviour of viscoelastic materials.

\begin{wrapfigure}[9]{l}{.3\columnwidth}
\centering
\vspace*{-0 cm}
\includegraphics[width = .3\columnwidth]{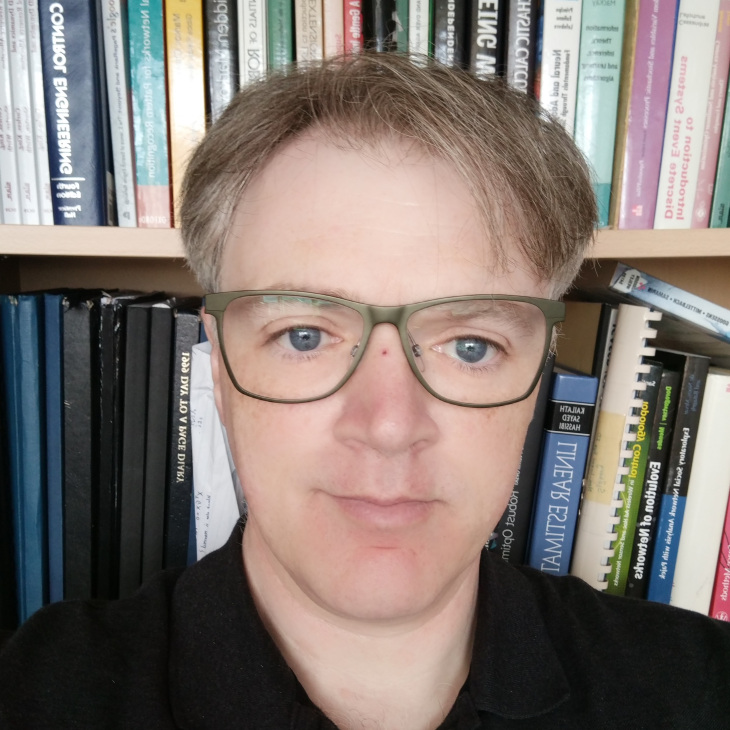}
\end{wrapfigure}
\textbf{Robert Shorten} is Professor of Cyber-physical Systems Design at Imperial College London and Professor of Control Engineering and Decision Science at UCD. He was a co-founder of the Hamilton Institute at Maynooth University, and led the Optimisation and Control team at IBM Research Smart Cities Research Lab in Dublin Ireland. He has been a visiting professor at TU Berlin and a research visitor at Yale University and Technion. He is the Irish member of the European Union Control Association assembly, a member of the IEEE Control Systems Society Technical Group on Smart Cities, and a member of the IFAC Technical Committees for Automotive Control, and for Discrete Event and Hybrid Systems. He is a co-author of the recently published books AIMD Dynamics and Distributed Resource Allocation (SIAM 2016) and Electric and Plug-in Vehicle Networks: Optimisation and Control (CRC Press, Taylor and Francis Group, 2017)

\end{document}